# Material Descriptors for the Discovery of Efficient Thermoelectrics

*Patrizio Graziosi,*[†‡] *Chathurangi Kumarasinghe,*[†] *Neophytos Neophytou*[†]

[†] School of Engineering, University of Warwick, Coventry, CV4 7AL, UK

[‡] Consiglio Nazionale delle Ricerche – Istituto per lo Studio dei Materiali Nanostrutturati, CNR – ISMN, via Gobetti 101, 40129, Bologna, Italy

ABSTRACT. The predictive performance screening of novel compounds can significantly promote the discovery of efficient, cheap, and non-toxic thermoelectric materials. Large efforts to implement machine-learning techniques coupled to materials databases are currently being undertaken, but the adopted computational methods can dramatically affect the outcome. With regards to electronic transport and power factor calculations, the most widely adopted and computationally efficient method, is the constant relaxation time approximation (CRT). This work goes beyond the CRT and adopts the proper, full energy and momentum dependencies of electron-phonon and ionized impurity scattering, to compute the electronic transport and perform power factor optimization for a group of half-Heusler alloys. Then the material parameters that determine the optimal power factor based on this more advanced treatment are identified. This enables the development of a set of significantly improved descriptors that can be used in materials screening studies, and which offer deeper insights into the underlying nature of high performance thermoelectric materials. We have identified $n_v \varepsilon_r / D_o^2 m_{\text{cond}}$ as the most useful and generic descriptor, a combination of the number of valleys, the dielectric constant, the conductivity effective mass, and the deformation potential for the dominant electron-phonon process. The proposed descriptors can accelerate the discovery of new efficient and environment friendly thermoelectric materials in a much more accurate and reliable manner, and some predictions for very high performance materials are presented.

1. Introduction

Thermoelectric generators (TEG) convert heat flow into useful electrical power and can provide energy savings, thus reduce our dependence on fossil fuels. However, the use of toxic materials without a sufficient energy conversion efficiency is currently hampering the large-scale exploitation of the thermoelectric (TE) technology. [1, 2, 3, 4] Efficient predictive simulation of materials' performance from first principles, by screening the myriad of possible compounds and alloys, can strongly boost the discovery of novel TE materials. [1, 2, 5-10] The accuracy and predicting ability of such efforts, however, strongly depend on the appropriateness of the screening parameters and descriptors for complex bandstructure materials, whereas the currently used descriptors are developed based on simplified parabolic bands and elastic phonon-limited scattering.[1] The proper identification of the parameters that govern the materials' performance, will accelerate the discovery process and drive the experimental research reliably.

The dimensionless figure of merit $ZT = \frac{\sigma S^2}{k_L + k_e} T$ quantifies the TE material conversion efficiency, where $\sigma$ is the electrical conductivity, $S$ the Seebeck coefficient, $k_L$ and $k_e$ are respectively the lattice and electronic thermal conductivities, and $T$ is the absolute temperature. [11] Materials research to date revolved around nanostructuring to reduce the thermal conductivity. [12, 13] Recently, several strategies to improve the power factor PF = $\sigma S^2$ have been identified, a direction which attracts significant interest as well. [14-17] Indeed, an analysis of the experimental literature on more than 11,000 compounds.[15, 18] reveals that the materials with the highest $ZT$ are generally found more often amongst the materials with high power factors.

Several material families are currently under intense investigation, such as clathrates, skutterudites, half-Heuslers, oxides, selenides, silicides, and others.[1, 17-19] Most of these materials have complex bandstructures, with multiple valleys and bands that extend in the entire Brillouin

Zone (BZ). In these materials, the electronic scattering processes also have complex energy, momentum and band dependencies. The most common computational approach is to extract the electronic bandstructures using *ab initio* methods, and afterwards use the Boltzmann Transport Equation (BTE) to extract the TE coefficients. [1, 20] As the extraction of the scattering rates used within the BTE is difficult, the constant relaxation time (CRT) or constant mean free path (CMFP) approximations are mostly used, where a CRT around 10 fs, or a CMFP of around 5 nm, are routinely employed. [5, 6, 10, 21, 22] These approximations have the advantage of being computationally efficient, but have the disadvantage of providing uncertain and rather arbitrary outcomes, both quantitatively and qualitatively, with respect to materials ranking, temperature trends, and optimal carrier density. [1, 23] Furthermore, by ignoring all energy, momentum and band dependence of the scattering processes, we have previously showed that the CRT approximation smears out the richness of the bandstructure that these complex materials possess. [23]

The importance of energy dependent scattering process treatment is gaining ground recently, as it is shown that accurately treating the energy dependencies can lead to much better, even opposite conclusions to what is predicted by simpler methods.[23] Computational methods vary in the treatment of electron-phonon scattering, depending on the required accuracy and computational complexities. In reference (24), by considering the energy dependence of acoustic phonon scattering, the TE coefficients were computed in the effective mass approximation and a combination of parameters to identify optimized materials was suggested. Reference (25) considered elastic scattering by acoustic phonons in a full band approach and identify the important role of the energy dependences, and the role of the (acoustic) deformation potential as a performance parameter. In references (2) and (20), the authors employed a full band approach and a numerical scheme for electron-phonon scattering based on calculating the electron-phonon coupling matrix elements. Due to the computational burden, however, it was proposed that a constant optical phonon frequency is used in the scattering description to sample the Brillouin zone.[2] Other than for 2D materials, it is computationally challenging to utilize the entire phonon and electron dispersions for full transport calculations, this, it is common to employ reasonable approximations.[20]

In this work, we perform an investigation for the thermoelectric power factor of a group of 14 half-Heusler compounds using DFT bandstructures and the full energy, momentum, and band dependence of electron-phonon scattering and ionized impurity scattering. We employ an advanced simulator with such capabilities, described in Section 2. We then compare, in section 3, the appropriateness of the currently used descriptors and develop a new set of descriptors that behave in a better way in predicting materials ranking in terms of the power factor, especially at the optimal doping level where the role of the ionized impurities is dominant. Finally, we discuss the impact of these descriptors on the understanding of the material properties that lead to optimal performance.

**2. Methods**

We use a novel simulator that considers all appropriate scattering mechanisms (acoustic phonon, optical phonon, ionized impurity scattering), and the full energy, momentum, and band dependence of the relaxation times.[23] The simulation approach consists of three stages: 1) calculation of the electronic structure using Density Functional Theory (DFT), 2) numerical extraction of the scattering rates, 3) calculation of the transport coefficients. The electronic bands are calculated using the Quantum Espresso package.[26, 27] The projector augmented wave technique is used with the PBE-GGA functional and the spin orbit coupling (SOC) has been considered as well. A kinetic energy cut-off greater than 60 Ry is used for the wave functions and an energy convergence criterion of $10^{-8}$ Ry for self-consistency was adopted. The computed bandgaps are very close to the experimental ones where available. For instance, for the NbFeSb compound we obtain a bandgap of 0.53 eV, in excellent agreement with the experimental value of 0.51 eV.[28]

The 3D bandstructure was calculated using a 51x51x51 Monkhorst–Pack $k$-point mesh on the primitive unit cell of the reciprocal lattice. The coordinates of the $k$-points, originally described in the reciprocal unit cell coordinate system, are transformed in the orthogonal Cartesian coordinate system. We then calculate the transport coefficient tensors in the Cartesian coordinates $x,y,z$ using the linearized Boltzmann Transport Equation (BTE). [16, 20, 29 - 31] The electrical conductivity $\sigma$ and the Seebeck coefficient $S$ tensors are expressed as

$$\sigma_{ij} = q_0^2 \int_E \Xi_{ij}(E)\left(-\frac{\partial f_0}{\partial E}\right) dE \quad (1a)$$

$$S_{ij} = \frac{q_0 k_B}{\sigma} \int_E \Xi_{ij}(E)\left(-\frac{\partial f_0}{\partial E}\right) \frac{E-E_F}{k_B T} dE \quad (1b)$$

where $E_F$, $T$, $q_0$, $k_B$, are the Fermi level, the absolute temperature, the electronic charge, and the Boltzmann constant, respectively, $f_0$ is the equilibrium Fermi distribution, and $ij$ are the Cartesian directions (in all the calculations we show, $i = j = x$). $\Xi(E)$ is the transport distribution function (TDF) defined as:

$$\Xi_{ij}(E) = \sum_{k,n}^{BZ} v_{i(k,n)} v_{j(k,n)} \tau_{i(k,n)} \delta_{(E_{k,n}-E)} = \frac{2}{(2\pi)^3} \oiint_{\mathfrak{L}_E^n} v_{i(k,n,E)} v_{j(k,n,E)} \tau_{i(k,n,E)} \frac{1}{\hbar} \frac{dA_{k,n,E}}{|v_{(k,n,E)}|} \quad (2)$$

where the sum over all the states in the Brillouin Zone (BZ) becomes a surface integral over the constant energy surface $\mathfrak{L}_E^n$ of energy $E$ defined by the delta function for the band of index $n$. In equation (2) $v_{i(k,n,E)}$ is the $i$-component of the band velocity of the transport state defined by the wave vector $k$ in the band $n$ at energy $E$, $\tau_{i(k,n,E)}$ is its momentum relaxation time (combining the relaxation times of each scattering mechanism, defined in equation (3) below, using Matthiessen's rule). $dA_{k,n,E}$ is the surface area element associated to the ($k,n,E$) transport state, $v_{(k,n,E)}$ is its band velocity and $\frac{1}{\hbar}\frac{dA_{k,n,E}}{|v_{(k,n,E)}|}$ its DOS.[23] For each transport state and each scattering mechanism $m_s$, the relaxation time $\tau_{i(k,n,E)}{}^{(m_s)}$ is given by a surface integral on the final constant energy surface $\mathfrak{L}_{E'}^{n'}$

$$\frac{1}{\tau_{i(k,n,E)}{}^{(m_s)}} = \frac{1}{(2\pi)^3} \oiint_{\mathfrak{L}_{E'}^{n'}} |S_{k,k'}{}^{(m_s)}| \frac{dA_{k'}}{\hbar|v_{(k')}|}\left(1 - \frac{v_{i(k')}}{v_{i(k,n,E)}}\right) \qquad (3)$$

where $k'$ represents a possible final state (the symbol $k'$ lumps its band index $n'$, that may be different from the initial one, and its energy $E'$, with $E' = E$ for elastic scattering and $E' = E \pm \hbar\omega$ in the case of inelastic scattering with phonon of frequency $\omega$).

$|S_{k,k'}|$ is the transition rate between $k$ and $k'$, and $v_i$ is the carrier velocity. The $\left(1 - \frac{v_{i(k')}}{v_{i(k,n,E)}}\right)$ term is an approximation for the momentum relaxation.[20, 23, 29 - 31] $|S_{k,k'}|$ is derived from Fermi's Golden Rule for the different scattering mechanisms,[23] and the relevant scattering parameters are taken from first-principle calculations.[9, 32, 33] Specifically, we employ deformation potential theory for electron-phonon scattering (acoustic and optical), and ionized impurity scattering with charge screening. [34] The corresponding transition rates for acoustic phonons, optical phonons and ionized impurity scattering are:

$$\left|S_{k,k'}{}^{(ADP)}\right| = \frac{\pi D_{ADP}^2 k_B T}{\hbar \rho v_S^2} \qquad (4a)$$

$$\left|S_{k,k'}{}^{(ODP)}\right| = \frac{\pi D_{ODP}^2}{\rho \omega}\left(N_{\omega,BE} + \frac{1}{2} \mp \frac{1}{2}\right) \qquad (4b)$$

$$\left|S_{k,k'}{}^{(IIS)}\right| = \frac{2\pi}{\hbar} \frac{Z^2 q_0^4}{\varepsilon_r^2 \varepsilon_0^2} \frac{N_{imp}}{\left(|k-k'|^2 + \frac{1}{L_D^2}\right)^2} \qquad (4c)$$

Above, $D_{ADP}$ is the acoustic deformation potential, $\rho$ is the mass density, $v_s$ is the sound velocity defined as $v_s = \frac{1}{3}s_l + \frac{2}{3}s_t$, which is important when the phonon bands are not isotropic, [23] where $s_l = \sqrt{\frac{K_V + 4/3 G_V}{\rho}}$ and $s_t = \sqrt{\frac{G_V}{\rho}}$ are the longitudinal and transverse sound speeds and $K_V$ and $G_V$ are the bulk and shear moduli. $D_{ODP}$ is the optical deformation potential, $\omega$ is the longitudinal optical phonon frequency in the single mode approximation with constant frequency over the entire reciprocal lattice unit cell. $N_\omega$ is its population density given by the Bose-Einstein statistics where the '+' and '−' signs indicate the absorption and emission processes, respectively. $N_{imp}$ is the impurity density, $Z$ the impurity charge, and $\varepsilon_0$ and $\varepsilon_r$ are the vacuum and the static relative permittivities. $L_D$ the Debye screening length in 3D defined as $L_D = \sqrt{\frac{\varepsilon_r \varepsilon_0}{q_0}\frac{\partial E_F}{\partial n}}$, where $n$ is the carrier density and $\partial n/\partial E_F$ is the variation of the carrier density with the Fermi level, which is temperature and doping dependent. The explicit use of $\partial n/\partial E_F$ enables us to apply the equation also under degenerate doping conditions. For the longitudinal optical phonon energy we adopt an averaged value that is assumed to be constant over the reciprocal lattice unit cell.[2, 23]

A representative example of the compounds studied is TiCoSb, whose bandstructure is shown in **Figure 1a**, and its conventional Zinc Blende unit cell in **Figure 1b**. Multiple bands of different curvatures in different directions compose the bandstructure and participate in transport. The shaded areas in **Figure 1a** highlight the energy range considered in our calculations, which extends up to 0.7 eV beyond the band edge, and corresponds to about $7k_B T$ beyond the highest Fermi level considered in the calculations at the highest temperature considered, $T = 900$ K. $7k_B T$ is largely sufficient as the largest part of the $\frac{\partial f_0}{\partial E}$ term in equation (1) extends for about $5k_B T$ above the Fermi level, and represents an optimal tradeoff between calculation accuracy and computational cost.[23] Examples of the carrier scattering transitions we consider in the calculations are sketched in **Figure 1c**, where two simple parabolic bands are displayed for illustrative purposes.

The selection rules and details of the strength of the electron-phonon coupling of each initial state to all other states individually are not yet well established for half-Heusler alloys in general. Thus, in this work we consider both intra-valley and inter-valley scattering events for phonon scattering, and allow transitions from an initial state to all other energy conserving states of the BZ. The Acoustic Deformation Potential (ADP) scattering that we consider is elastic. The scattering with optical phonons of energy $\hbar\omega$, considered within the Optical Deformation Potential (ODP) theory, is inelastic. Phonon scattering is considered both intra and inter-band. The scattering due to ionized dopant impurities, (IIS), is elastic and considered as intra-band only, following the common treatment in the literature.[30, 35, 36] We adopt the relative dielectric constants from reference (9) where they were calculated from first principles. The full list of first principles computed input parameters is listed in the Supporting Information.

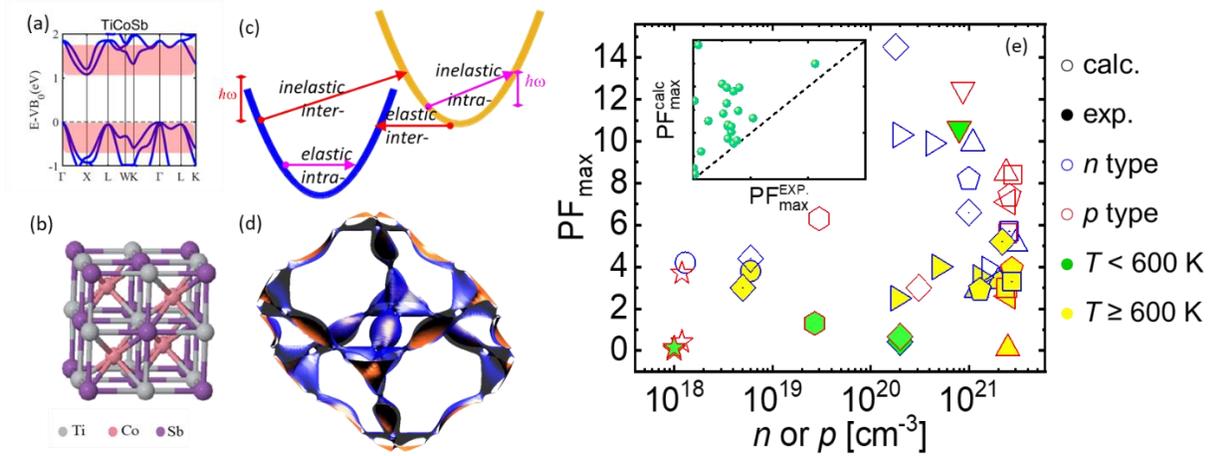

**Figure 1.** (a) Bandstructure of TiCoSb, as a representative example. The shaded regions highlight the energy windows used in the calculations. (b) Conventional cubic zincblende unit cell of TiCoSb as a representative of all 14 compounds investigated. (c) Types of scattering transition we consider depicted using two parabolic bands. The inelastic processes are due to the absorption or emission of optical phonons of energy $\hbar\omega$. (d) Reciprocal space constant energy surfaces at $E = -0.125$ eV for two bands (blue and orange) below the valence band edge of TiCoSb. Warped shapes with narrow wings and tubular connections between minima can be observed. (e) Experimental (filled symbols) and calculated (open symbols) power factor versus carrier concentration. Symbols with blue edge are for *n*-type materials and with red edge are for *p*-type materials. The filling color is an indication of the temperature: green for $T < 600$ K and yellow for $T \geq 600$ K. Each material is indicated by the symbols: HfCoSb: square,[37] HfNiSn: circle,[38] NbCoSn: up-pointing triangle,[39, 40] NbFeSb: down-pointing triangle, [28] ScNiSb: diamond,[41] TiCoSb: left-pointing triangle,[37] TiNiSn: right-pointing triangle,[42, 43] YNiBi: hexagon,[44] YNiSb: star,[41] ZrCoBi: pentagon,[45] ZrCoSb: dot centered square,[37, 45, 46] ZrNiSn: dot centered diamond.[47, 48] The inset compares the computed and measured PFs at similar conditions related to the carrier density and temperature conditions. The dashed line has unity slope.

The TDF defined in equation (2) is a surface integral on the relevant constant energy surfaces. An example of such surfaces is displayed in **Figure 1d**, where two constant energy surfaces from two valence bands of TiCoSb at 0.125 eV below the band edge are displayed (blue and orange). These surfaces are warped, with elongated regions, necks and pockets, very different from the regular ellipsoids that enable analytical expressions for the relaxation times. They extend over nearly the whole reciprocal unit cell, all of which needs then to be included numerically in all scattering events. A full band numerical treatment becomes compulsory, which is done by the extraction of the k-points at constant energies, essentially converting the usual E($k$) into a k(E).[23] The transport calculations presented here can require from ~ 20 up to ~ 600 computation CPU hours for each material, depending on the bandstructure complexity, executed in parallel on ~ 20 CPUs.

3. Results and discussion

This section presents the main results, starting from the transport properties of the half-Heuslers considered, to the developed material descriptors for transport under: i) phonon-limited and then under phonon plus IIS transport, ii) at 300 K first and then at higher temperatures, and iii) under unipolar and bipolar transport.

An initial glance at the results is shown in **Figure 1e**, that presents measured PF values versus carrier density, and compares them with the computed values at the same (or nearby) carrier density and temperature. [28, 37 - 48] The symbols, one for each material, denote: filled/empty for experimental/calculated, blue/red boundary for *n*-/*p*-type, and green/yellow filling for measured temperature below/above 600 K. The inset shows the correlation between computed and experimental values, with the dashed line having unity slope. In some cases, the calculation shows excellent agreement with experimental values, but in most cases the calculated PFs are higher than the measured ones by a factor of two or three. Considering that experimental samples are polycrystalline while our calculations assume perfect single crystals, the similarity in the PF values indicates the credibility of our computational method. [23] An interesting observation is the case of *p*-type NbFeSb (down-pointing triangles at PF ∼ 10 mW/mK$^2$ and densities ∼ 10$^{21}$ cm$^{-3}$). The computed PF values are very close to the measured ones, and those samples have the highest degree of crystallinity, with the largest grain sizes.[28]

### 3.1 Thermoelectric charge transport properties

Both electron transport for *n*-type materials and hole transport for *p*-type materials are considered. We compute the conductivity $\sigma$, the Seebeck coefficient $S$, and the power factor PF = $\sigma S^2$, under three scattering scenarios: i) phonon-limited transport $\tau_{ph}(E)$, ii) scattering of charge carriers with both phonons and ionized impurities $\tau_{ph,IIS}(E)$, and iii) CRT with $\tau_c$ = 10 fs, as commonly adopted in the literature.[1, 5, 7, 10, 21] The conduction and valence bands are analyzed separately assuming unipolar transport, but also combined later on, in section 3.3, to describe bipolar transport. We plot all quantities in terms of the reduced Femi level $\eta_F$, which denotes the distance of the Fermi level from the band edge, and has a one-to-one correspondence with the carrier density.

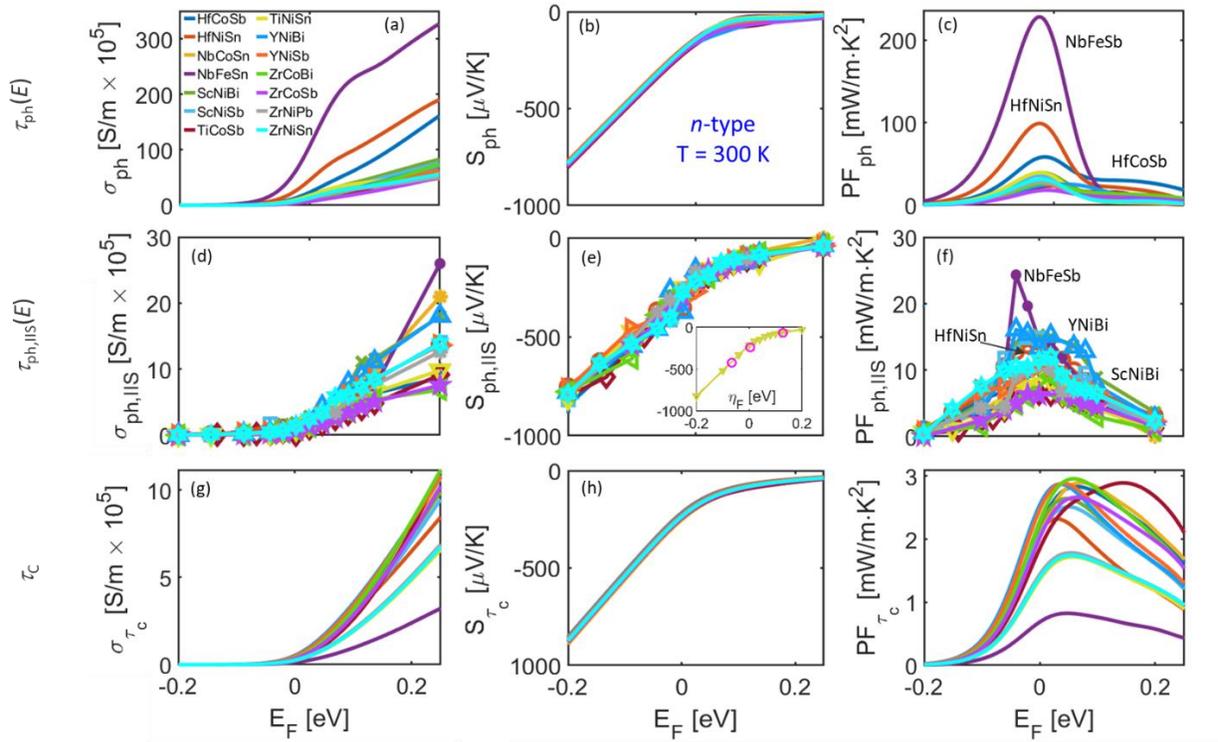

**Figure 2.** *n*-type thermoelectric coefficients versus the relative position of the Fermi level $\eta_F$ for the 14 half-Heusler compounds under investigation for unipolar transport. Row-wise, simulation results for three different scattering scenarios are presented: (a-c) electron phonon-limited scattering $\tau_{ph}(E)$; (d-f) phonon plus ionized impurity scattering, $\tau_{ph,IIS}(E)$; (g-i) constant relaxation

time (CRT) approximation with $\tau_c$ = 10 fs. Column-wise: electrical conductivity $\sigma$ (a), (d) and (g), Seebeck coefficient $S$ (b), (e), (h), and power factor PF (c), (f), (i). The data refer to room temperature conditions. The inset in (e) compares the computed Seebeck coefficients for TiNiSn (dark yellow) with the corresponding experimental values (magenta dots) measured at room temperature in *quasi*-single crystals.[49]

### 3.1.1 *n*-type thermoelectric coefficients (unipolar transport)

The dependence of the charge transport coefficients ($\sigma$, $S$ and PF) of the considered half-Heuslers upon $\eta_F$ at room temperature ($T$ = 300 K) is shown in **Figure 2**. $\eta_F$ = 0 means that the Fermi level is at the band onset, $\eta_F$ < 0 in the band gap, and $\eta_F$ > 0 is for heavily doped, degenerate conditions with the Fermi level into the conduction band. Each compound is depicted by a specific color as displayed in the legend in **Figure 2a**. The transport coefficients $\sigma$, $S$ and PF are displayed column-wise, while the different rows correspond to the different scattering scenarios: scattering with phonons only $\tau_{ph}$ ($E$), the adding ionized impurity scattering (IIS) $\tau_{ph,IIS}$ ($E$), and the CRT approximation $\tau_c$.

The electrical conductivity, $\sigma$, in **Figures 2a, 2d, 2g** (first column), displays a large variation between the materials, while the $S$ values, in **Figures 2b, 2e, 2h** (second column), are similar for all the materials for a given scattering scenario. Thus, the electrical conductivity drives the materials ranking, and it is also the quantity that is most sensitive to the specific details used to compute the relaxation times.

We note that in the CRT approximation the Seebeck coefficient $S$ does not depend on the value of the relaxation time, and it is in general considered weakly sensitive to the scattering details as well. However, when we consider the energy/wave-vector dependence of the relaxation time, we observe somewhat quantitatively different results compared to the CRT approximation. The average Seebeck coefficient between materials at $\eta_F = 0$ in the phonon-limited case is ~ 250 µV/K, in the case of phonon plus IIS ranges between 200 and 300 µV/K, but in the CRT case it is ~ 300 µV/K. Under IIS consideration, a larger variation between the Seebeck coefficients of the materials is observed (in the range of ~ 50 % around $\eta_F = 0$), whereas in the $\tau_{ph}(E)$ and CRT cases the variation is only ~ 10 %. Thus, a proper scattering treatment, including both, energy dependencies and ionized impurity scattering, should be conducted also in the evaluation of the Seebeck coefficient, rather than only in the case of the conductivity. The inset of **Figure 2e** compares the *S* values computed for TiNiSn with experimental values in *quasi*-single crystals (magenta points).[49] We use values of $\eta_F$ which map the experimental carrier density. The excellent agreement supports our computational approach.

The ranking of materials with respect to the optimal PF (third column, **Figures 2c, 2f** and **2i**), varies significantly under the different scattering scenarios, driven by the variations in $\sigma$. The optimal Fermi level is at the band edge in the case of phonon-limited scattering ($\eta_F = 0$ in Figure 2c). When IIS is also considered in **Figure 2f**, the ideal Fermi level position moves from few tens of meV in the bandgap (negative $\eta_F$, extreme case for NbFeSb with optimal $\eta_F$ around -40 meV) to few meV into the band (positive $\eta_F$, extreme case of NbCoSn with optimal $\eta_F$ around 10 meV). On the contrary, the CRT approximation suggests degenerate conditions as being the optimal ones for all the materials, suggesting even up to one order of magnitude higher doping, [23] which is also more difficult to achieve experimentally.

The best candidates for *n*-type TE materials are NbFeSb (purple line) for its low deformation potentials ($D_{ADP} = 1.0$ eV, $D_{ODP} = 1.6 \times 10^{10}$ eV/m), low conductivity effective mass $m_{cond}$ (0.33 $m_0$), and high dielectric constant ($\varepsilon_r = 23.0$). We refer to $m_{cond}$ as the effective mass of an isotropic parabolic band that gives the same carrier velocity as the overall material's full bandstructure, which is a quantity experimentally accessible; [50] see Supporting Information for more details on how we extract it. HfCoSb and HfNiSn also have quite low deformation potentials ($D_{ADP}$ = 0.2 eV, $D_{ODP}$ = 1.4×10$^{10}$ eV/m for HfCoSb and $D_{ADP}$ = 0.1 eV, $D_{ODP}$ = 1.8×10$^{10}$ eV/m for HfNiSn), but relatively lower $\varepsilon_r$ (especially HfCoSb, $\varepsilon_r$ = 17.5). The ranking of these materials slightly changes when IIS is added, possibly because the larger $\varepsilon_r$ increases the charge screening and improves the conductivity (as we show below). Vice versa, the high $\varepsilon_r$ (26.6, 20.8, 29.0) could favor YNiBi, HfNiSn, ScNiBi, respectively, which end up in second place after NbFeSb. It is worth mentioning that from those only HfNiSn is a known *n*-type half-Heusler with Sb as the donor in the lattice place of Sn while the other compounds are usually p-type.[51,52] Notice that the advantage of NbFeSb is lost in the case of the CRT approximation, which predicts that NbFeSb is the worst performer.

3.1.2 *p*-type thermoelectric coefficients (unipolar transport)

The same analysis is now applied to hole transport for the same compounds. The dependence of the charge transport coefficients ($\sigma$, $S$ and PF) for different Fermi level positions at 300 K is displayed in **Figure 3**, as above in **Figure 2**. Column-wise we show $\sigma$, $S$ and PF, and row-wise the three different scattering scenarios. In the case of hole transport, $\eta_F > 0$ describes a Fermi level residing in the bandgap and $\eta_F < 0$ indicates heavily doped conditions with the Fermi level into the valence band. Each compound is depicted by the same color as in **Figure 2**. As in the case of *n*-type transport, the electrical conductivity in the first column, **Figures 3a, 3e, 3g**, is the transport coefficient that displays the larger variation between materials and drives the materials ranking.

Again, the optimal doping appears when the Fermi level is placed around the band edge when energy dependent scattering physics is considered ($\eta_F$ = 0 in **Figure 3c**, and **3f**). The only exceptions are ScNiBi and ScNiSb, for which the PFs peak at a Fermi level slightly in the gap. Under the CRT approximation, in **Figure 3i**, the PFs peak at heavily doped degenerate conditions for all the materials, with Fermi levels positioned from 0.05 eV to 0.1 eV into the band, suggesting an optimal carrier density around one order of magnitude higher.[23]

The best candidates for *p*-type TE materials are ScNiBi for its low deformation potentials ($D_{ADP}$ = 0.2 eV, $D_{ODP}$ = 1.4×10$^{10}$ eV/m), low $m_{cond}$ (0.44 $m_0$), and very high dielectric constant ($\varepsilon_r$ = 29.0). TiNiSn is a top performer under phonon-limited considerations for its low deformation potentials ($D_{ADP}$ = 0.2 eV, $D_{ODP}$ = 0.9 × 10$^{10}$ eV/m). Fortunately, ScNiBi can be *p*-doped when substituting Bi with Sn[52] and NbCoSn and NbFeSb can be *p*-doped when replacing Nb with Ti or Zr,[39] or by substitutional doping with Fe in the Co site, like in TiCoSb.[53] When it comes to the CRT approximation, only the NbCoSn (orange line) continues to perform well, but ScNiBi and TiNiSn lose their advantage. The reason NbCoSn still performs well is its simultaneous small conductivity effective mass and high DOS effective mass, as we will highlight later on in the development of the material descriptors.

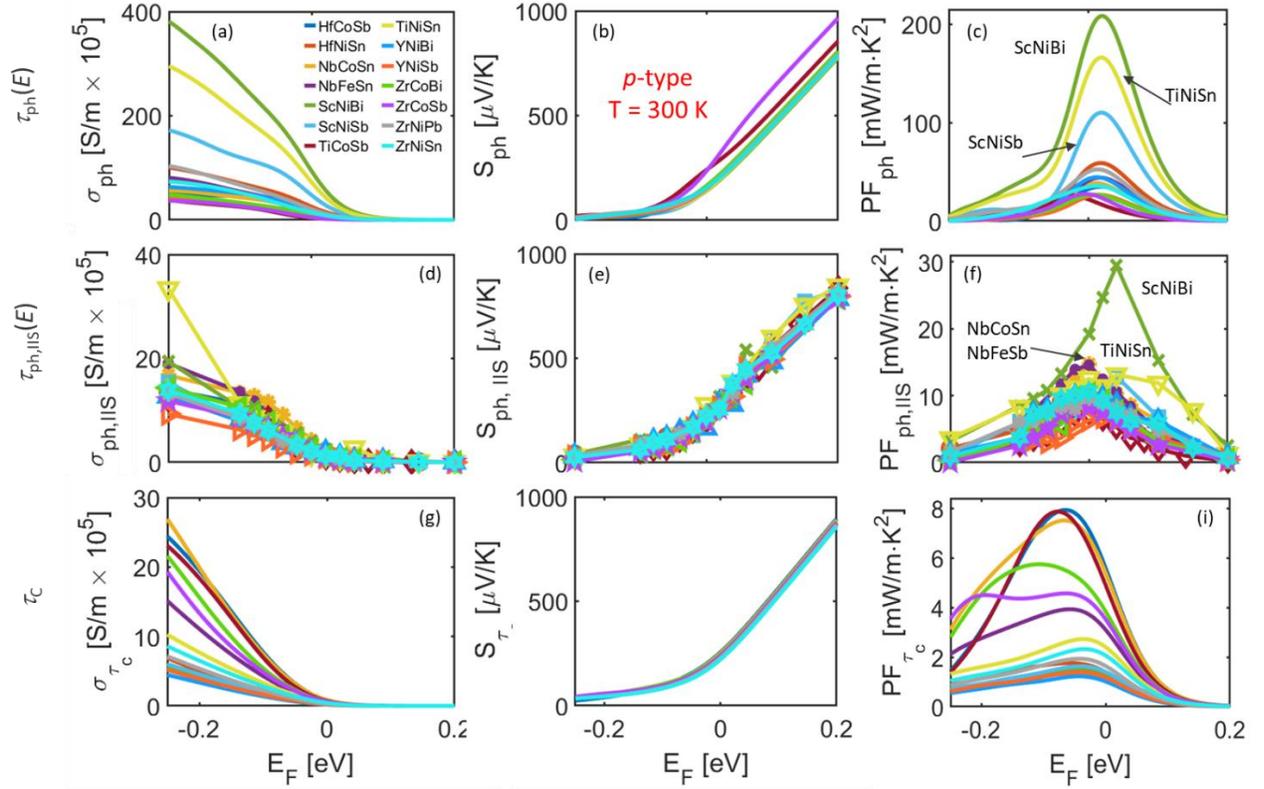

**Figure 3.** p-type thermoelectric coefficients versus the relative position of the Fermi level $\eta_F$ for the 14 half-Heusler compounds under investigation for unipolar transport. Row-wise, simulation results for three different scattering scenarios are presented: (a-c) electron phonon-limited scattering $\tau_{ph}(E)$; (d-f) phonon plus ionized impurity scattering, $\tau_{ph,IIS}(E)$; (g-i) constant relaxation time (CRT) approximation with $\tau_c = 10$ fs. Column-wise: Electrical conductivity $\sigma$ (a), (d) and (g), Seebeck coefficient S (b), (e), (h), and power factor PF (c), (f), (i).

Importantly, however, each scattering scenario can result in different predictions for the PF. In **Figure 4**, for example, we show the correlation between the PF values computed under the three different scattering assumptions. Blue dots are for electron transport data and red dots for hole transport data. In the legend, the $r_{e/h}$ are the Pearson correlation coefficients for electron/hole transport, that range from 1 (complete correlation) to -1 (complete anticorrelation). **Figure 4a** shows the correlation between the peak PF values computed under the CRT approximation ($\tau_c$) and under phonon-limited scattering treatment, $\tau_{ph}(E)$. The ranking data are largely uncorrelated, even nearly anticorrelated in the *n*-type case, which reflects back to the very different materials ranking observed in the two scattering cases. A similar absence of correlation between the PF computed values still holds when comparing the CRT results with the phonon plus IIS scattering, $\tau_{ph,IIS}(E)$, shown in **Figure 4b**. On the other hand, comparing the computed PF under the $\tau_{ph}(E)$ and the $\tau_{ph,IIS}(E)$ cases, in **Figure 4c**, a sufficient correlation appears for both the *n*-type and *p*-type materials ($r_e = 0.82$ and $r_h = 0.79$).

**3.2 Descriptors development**

3.2.1   Available Descriptors

Large efforts are currently being undertaken in order to identify potential high performance TE materials and their alloys using materials screening methods, guided by machine learning techniques.[1, 54] A tremendously large set of materials data is accumulated in databases.[18, 55-58] Such efforts define appropriate descriptors, i.e. simple parameters or a combination of them, which could signify high TE performance, for example the number of valleys, density of states, effective mass, mobility, etc., and then the material databases are screened to find materials that maximize the descriptor values. In this way, high-throughput screening can be performed, but the confidence in the descriptors needs to be high.

A widely used descriptor is the dimensionless quality factor *β*, or "B-factor",[1, 11] defined as $\beta = n_v \mu_v \frac{T k_B^2}{e \kappa_L}$, where $n_v$ is the number of valleys, $\mu_v$ is the mobility of the carriers in a single valley, $\kappa_L$ is the lattice thermal conductivity, *T* is the temperature, $k_B$ the Boltzmann constant, and $e$ the electron charge.[11] A slight variation of this is $\frac{\mu_0 m_{DOS}^{*3/2}}{\kappa_L}$, where $\mu_0$ is the mobility and $m_{DOS}^*$ is the DOS effective mass.[1] The numerators, $n_v\mu_v$, or $\mu_0 m_{DOS}^{*3/2}$, are related to the power factor. Another variation, developed from the mobility dependence on acoustic phonon-limited scattering, is $\frac{n_v c_l}{m_v D_{ac}^2}$, where $n_v$ is the number of valleys, $m_v$ the density of states (DOS) effective mass for a single valley, $D_{ac}$ the deformation potential for acoustic phonons, $c_l = \rho u_s^2$ with $\rho$ the material density and $u_s$ the sound velocity.[59] All these descriptors originate from a semi-empirical description of charge transport under the simple assumptions that the bands are parabolic and the charge transport is phonon-limited.[1, 11]

Recent studies that use full numerical bandstructures coupled to the extraction of the electron-phonon scattering rate via the direct calculation of the interaction matrix elements, suggested the inverse of the density of states effective mass, $1/m_{DOS}^*$, as a single-parameter descriptor.[2] While this is a valuable outcome, it is still based on phonon-limited scattering conditions.[1]

The weak correlation in the materials ranking between each scattering scenario suggests that each case is optimized by different descriptors, which points to the necessity of accurate descriptors to obtain any meaningful results in materials screening. Under the CRT approximation, a descriptor that offers satisfactory Pearson coefficients for both carrier types is $m_{DOS}^{3/2}/m_{cond}$, as shown in **Figure 4d**. The numerator directly relates to the DOS and the denominator to the squared velocity of the carriers, as it appears in the TDF, equation (2). Here, the $m_{DOS}$ is the density of states effective mass (the effective mass of an isotropic parabolic band that gives the same DOS as the actual bandstructure at the band onset), and $m_{cond}$ is the conductivity effective mass, which is related to the carrier velocity. Details on the extraction of these quantities can be found in the Supporting Information. These parameters are related to the whole bandstructure and automatically include the effect of multiple, and different, valleys. Such descriptor points reasonably towards seeking materials with many and fast carriers, and indicates that the CRT approximation favors materials with higher DOS effective mass.

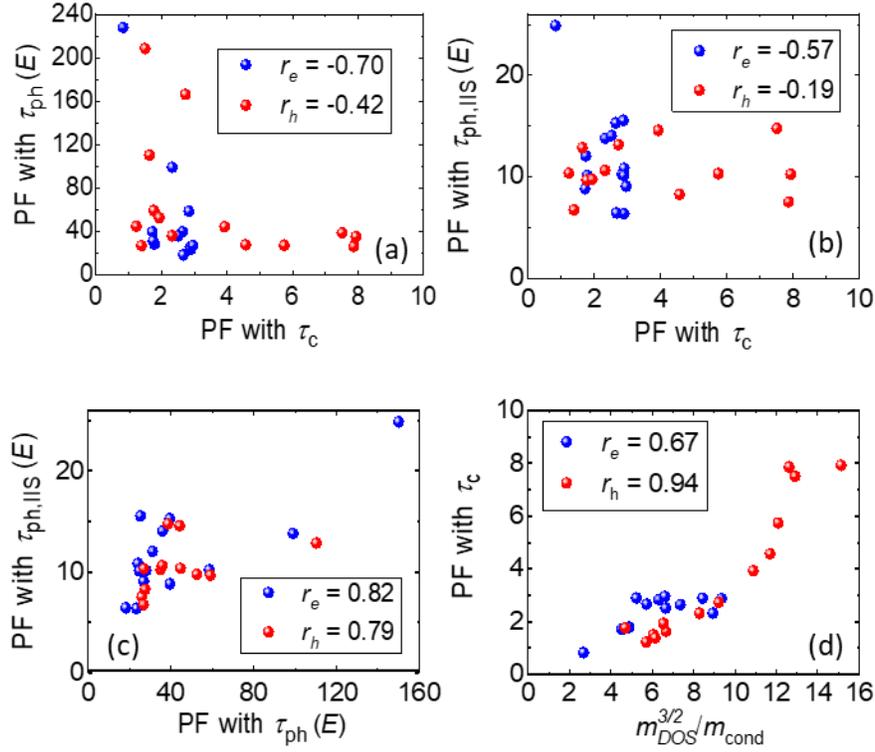

**Figure 4.** Correlation between the peak PF values computed under different scattering scenarios: constant relaxation time $\tau_c$, electron phonon scattering $\tau_{ph}(E)$, and electron-phonon plus ionized impurity scattering $\tau_{ph,IIS}(E)$. The blue and red colors are for the *n*-type and *p*-type materials, respectively. (a) Correlation between CRT and $\tau_{ph}(E)$ transport; (b) correlation between CRT and $\tau_{ph,IIS}(E)$ transport; (c) correlation between $\tau_{ph}(E)$ and $\tau_{ph,IIS}(E)$ transport. (d) A descriptor proposed in this work for the CRT approximation: the ratio between DOS effective mass ($m_{DOS}$) raised to the 3/2 power over the conductivity effective mass $m_{cond}$. The Pearson correlation coefficient $r$ is displayed in the legend, $r_e$ for electron transport and $r_h$ for hole transport.

The situation is less straightforward when we consider energy-dependent electron-phonon scattering, because in addition to having more conducting states, a higher DOS results in a stronger scattering rate, [23, 60] and the two effects could cancel each other out.[23, 27, 60] We first examine the adequacy of existing descriptors already proposed in the literature, in **Figure 5**. We begin with the commonly used $\frac{n_v c_l}{m_v D_{ac}^2}$, in **Figure 5a**. [59] $m_v$ represents an 'average' DOS mass of a single valley such that if its density of states is multiplied by the number of valleys, the overall bandstructure DOS is obtained. This is given by $m_{\text{DOS}/1v} = (n_v)^{-\frac{2}{3}} m_{\text{DOS}}$ that gives a total DOS same as the sum of the DOS of the individual valleys.[34] The unsatisfactory performance of this descriptor with respect to our calculations ($r_e$ = -0.04 and $r_h$ = 0.04) resides in two reasons. First, the descriptor relies on intra-valley scattering alone, while here we consider inter-valley/inter-band scattering as well. In the case of intra-valley scattering only and phonon-limited transport, the quality factor is likely to provide better correlation. However, in the case of inter-band/inter-valley transitions, the increased scattering into the larger DOS of the final scattering states cancels out the potential benefit, which removes the influence of the number of valleys and effective mass. [23, 60] In general, whether intra- versus inter-band scattering dominates is not known for these materials, but we have no reason to neglect the inter-band/inter-valley, especially when the dominant phonons are the optical ones. Second, acoustic phonon deformation potential in the half-Heuslers we examine is quite low compared to the optical phonon deformation potential which dominates, [9] which is why we use it in the formation of the descriptor. Indeed, the band edge states in the conduction and valence bands of half-Heuslers, are characterized by nearly non-bonding orbitals. [9] This is a situation under which the bonding/anti-bonding interactions of the orbitals that form the band edge states nearly cancel, making the band edges resilient to energy shifts upon structure deformations, thus providing weak electron-acoustic phonon interactions. This resilience is enhanced by the fact that the interactions of those orbitals with orbitals forming other energy states in nearby or higher energies are mostly symmetry forbidden. This does not happen for the electron-optical phonon interactions, which then acquire a significantly higher deformation potential. Indeed, upon substitution of the acoustic deformation potential $D_{ac}$ with the optical phonon one, $D_o$, in **Figure 5b**, much better agreement is achieved for the *n*-type case ($r_e$ = 0.72), but still not for the *p*-type. The reason is that all *n*-type materials we consider have the same number of valleys (only X valleys). Thus, whether intra- or inter-valley scattering is considered, it affects all of them similarly. That is not the case for the *p*-type materials, which have several and varying number of valleys that enter the descriptor (see the Supporting Information for details), but drop out of the TDF in equation (2).

Thus, the quality factor descriptor provides adequate correlation for *n*-type when it includes the dominant deformation potential, regardless it being $D_{ac}$ or $D_o$. Experimentally, the deformation potentials are extracted from the temperature dependence of the mobility, and by assuming a certain effective mass value. Under the effective mass approximation, under either ADP or ODP, the same mobility temperature dependence is achieved.[34,61] Thus, one assumes that the extracted deformation potential will somehow effectively contain both.[61]

The single parameter descriptor $1/m_{DOS}$,[2] is displayed in **Figure 5c** and performs well for electron transport ($r_e$ = 0.72) but not for hole transport ($r_h$ = 0.25). This is consistent with reference (2), in which the relaxation time versus the $m_{DOS}$ trend is clearly linear for *n*-type half-Heusler alloys, but not for the *p*-type case.

**Figures 5d, 5e** and **5f** show the correlation between the peak PF computed at 300 K and the same descriptors as in **Figures 5a, 5b and 5c**, when we further include the IIS mechanism, $\tau_{ph,IIS}(E)$. This comparison is important, because this is the default experimental situation. The quality factor,[59] **Figure 5d**, and its adjustment with $D_o$ instead of $D_{ac}$, **Figure 5e**, still provides acceptable Pearson

correlation coefficients for electrons ($r_e = 0.60$), which somehow justifies its use in rationalizing experimental measurements. [45,59] As for the phonon-limited case, the $1/m_{DOS}$ descriptor works much better for electrons than for holes ($r_e = 0.76$, $r_h = 0.19$), **Figure 5f**. This is not surprising, as this descriptor is specifically suggested from phonon-limited scattering considerations.

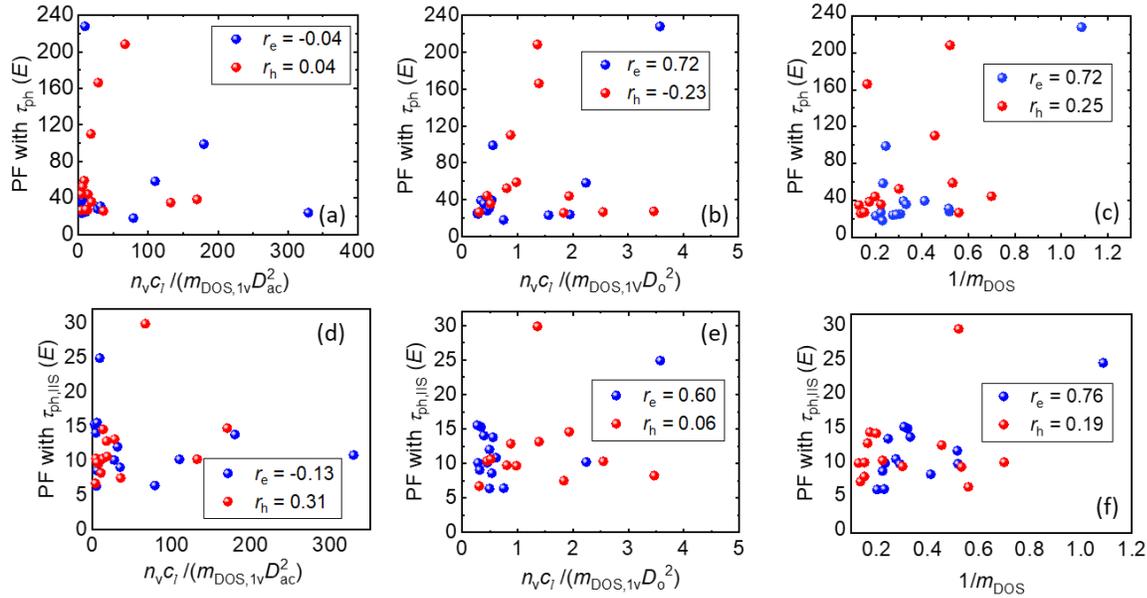

**Figure 5.** Correlation between the peak PF and common descriptors proposed in the literature for the phonon-limited transport, $\tau_{ph}(E)$, top row, and the phonon plus ionized impurity scattering transport, $\tau_{ph,IIS}(E)$, bottom row. The Pearson correlation coefficient $r$ is displayed in the legends. (a), (d) A descriptor based on the single band mobility and the number of valleys proposed by Tang *et al.* [59] (b), (e) The same as in (a) and (d) but with the acoustic deformation potential $D_{ac}$, substituted by the non-polar optical deformation potential $D_o$ that is dominant in these materials. (c), (f) The DOS effective mass $m_{DOS}$ descriptor. [2]

### 3.2.2 Development of new descriptors

After, testing the existing descriptors, we present here new descriptors based on the energy and wave-vector dependent scattering mechanisms, as well as IIS, which would be more appropriate for materials screening. The descriptor development is depicted in **Figure 6**. We first examine the phonon-limited scattering case (top row) and then the phonon plus IIS case (bottom row). In the search for a single parameter descriptor, the ODP deformation potential $D_o$ represents a first intuitive guess because this inelastic scattering mechanism has much higher strength amongst these compounds compared to $D_{ac}$.[10] The correlation between the peak PF values and $1/D_o^2$ in **Figure 6a** is fairly satisfactory for both *n*-type and *p*-type cases ($r_e$ = 0.54, $r_h$ = 0.69). The fact that the scattering strength is proportional to the square of the deformation potential motivates the choice of squared $D_o$. This descriptor can be further improved by carrier velocity considerations, by including the bandstructure conductivity effective mass $m_{\text{cond}}$. The $1/D_o^2\, m_{\text{cond}}$ descriptor in **Figure 6b** offers an astonishing agreement ($r_e$ = 0.96, $r_h$ = 0.95).

We deliberately avoid using parameters related to the DOS due to the reason that under energy dependent scattering the DOS cancels out with the inverse DOS dependence of the relaxation times, and at first order drops out of the transport coefficients.[23, 60] Thus, when looking for effective mass parameters, we chose the conductivity effective mass, which does not cancel out in the evaluation of the TDF in equation (2). Since the deformation potential and conductivity effective mass are two parameters related to the mobility, **Figure 6c** plots the peak PF values versus the computed lattice mobility (low field and low carrier density), a measurable quantity related to electron-phonon scattering processes. We do not choose the mobility at the carrier density of the PF peak, because the mobility of an experimental specimen at these carrier densities is limited by the IIS. The mobility correlation is very good for *n*-type materials ($r_e$ = 0.96) and fairly good for *p*-type ($r_h$ =

0.78) as also suggested in the past.[1,11] The mobility can be a valid single parameter descriptor for machine learning studies in databases that collect experimental results,[18] but in the case of computational studies it is not readily available.

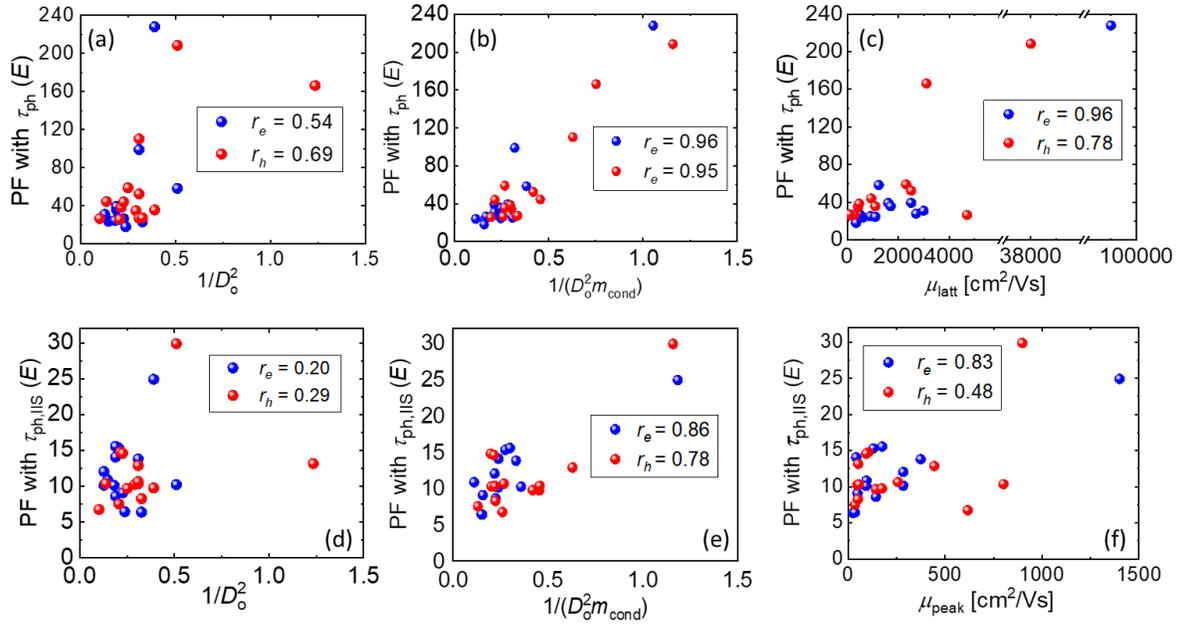

**Figure 6.** Correlations between the peak of the PF and various performance descriptors proposed in this work for phonon-limited transport, $\tau_{ph}(E)$, first row, and phonon plus ionized impurity scattering transport, $\tau_{ph,IIS}(E)$, second row. (a) A single parameter descriptor based on the optical deformation potential $D_o$. (b) The descriptor in (a) is improved by including the conductivity effective mass $m_{cond}$. (c) The correlation of the PF peak with the computed lattice low-field, low-density mobility, which is a parameter accessible in experiments. (d-f) Same as the descriptors in (a-c) but applied to the peak PF values computed when also including the IIS mechanism. The computed mobility in (f) is at the density where the peak PF is achieved for each material and carrier type, a parameter accessible in experiments.

Adding ionized impurity scattering (IIS) to electron-phonon scattering, $\tau_{ph,IIS}(E)$, comprises a picture closer to realistic TE materials at the doping densities of interest. Testing the descriptors of **Figure 6a**, **6b** and **6c** for the peak of the PFs in the phonon plus IIS case, though, indicates reduced correlation. The $1/D_o^2$ descriptor in **Figure 6d** is inadequate for both *n*-type ($r_e = 0.20$) and *p*-type cases ($r_h = 0.29$). The addition of the conductivity effective mass ($1/D_o^2 m_{cond}$) in **Figure 6e** largely improves the correlation for both *n*-type ($r_e = 0.86$) and *p*-type materials ($r_h = 0.78$). In addition, the mobility parameter in **Figure 6f** (calculated at the density where the peak of the PF appears), which is useful from an experimental point of view, seems applicable for *n*-type half-Heusler TE materials ($r_e = 0.83$) but again not for the p-type materials ($r_h = 0.48$). Thus, in the presence of the dominant IIS, $1/D_o^2 m_{cond}$ seems to be a satisfactory descriptor for the PF in the nearly parabolic conduction band of the considered half-Heuslers, but it is not adequate for the strongly warped, non-parabolic bandstructures like the valence bands of these materials, in contrast to phonon-limited transport conditions. Thus, a further improvement to the descriptor is necessary.

The material dependent term that appears in the IIS scattering rate, is proportional to $\frac{1}{\varepsilon_r^2}\left(\frac{1}{q^2 + 1/L_D^2}\right)^2$, with $\varepsilon_r$ being the relative permittivity, $L_D$ the Debye screening length and $q^2 = |k - k'|^2$ the scattering vector between initial state $k$ and the final state $k'$.[23] $L_D = \sqrt{\frac{\varepsilon_r \varepsilon_0}{e}\left(\frac{\partial n}{\partial E_F}\right)^{-1}}$ is the screening length, with $\varepsilon_0$ the vacuum permittivity, $e$ the electronic charge, $E_F$ the Fermi level and $n$ the carrier density. That term is then proportional to $\left(\frac{1}{q^2 L_D^2 + 1}\right)^2$, in which case the material-dependent parameters appear through the $q^2 L_D^2 \sim q^2 \varepsilon_r$ term, and they become important if it is bigger than the unity. The screening lengths are around $\sim 1$ nm ($L_D$ at the PF peak ranges from 0.3 nm to 1.9 nm depending on the material). With the cubic lattice constant being around $\sim 0.6$ nm, the $q^2\varepsilon_r$ term becomes important in suppressing scattering and increasing the PF for transitions for which the states are farther than 15% (from 5% to 30% depending on the material) of the Brillouin Zone. This essentially creates conditions for inter-valleys scattering suppression for IIS.

We developed a method to estimate an average $\langle q \rangle$ to improve the descriptor. For each possible pair of initial and final states we compute the scattering vector (**Figure 7a**) and average it for all the *k*-points of the constant energy surface *E* and band index *n* to get a $\langle q \rangle_{E,n}$. Then, we perform the energy average using the density of states $g_{(E)}$ and the Fermi distribution $f_{(E)}$ centered at the band edge, where the peak PF occurs as $\langle q \rangle_n = \frac{\sum_E \langle q \rangle_{n,E} f_{(E)} g_{(E)}}{\sum_E f_{(E)} g_{(E)}}$. Finally, we compute $\langle q \rangle$ by averaging the $\langle q \rangle_n$ over the bands.

The correlation between the peak PF and the descriptor containing $q^2\varepsilon_r$, as $\langle q \rangle^2 \varepsilon_r / D_o^2 m_{\text{cond}}$, is shown in **Figure 7b**, with $r_e = 0.88$ and $r_h = 0.82$. Effectively, this indicates that to achieve high correlation for *p*-type materials, the descriptor needs to include features of the IIS screening details, which reflect on the size of the constant energy surfaces. In bands with many valleys separated by low energy maxima, the constant energy surfaces can be very large and connect points very far in the BZ, even for energies not so far from the band edge. In this situation, the scattering vector can be very large. When the scattering vector becomes very large, it can dominate screening and play an important role in conductivity, the PF, and materials' ranking.

The extraction of such average scattering vector, however, might not be straightforward, and it certainly defeats the purpose of having descriptors that are easily constructed on simple parameters. A simple way to capture the size of the energy surface is to consider the number of valleys in each band. The larger the number of valleys (from the same band), the wider the surfaces are. For this, we count the number of valleys in a range of 0.1 eV from the band edge. Essentially, this says that large constant energy surfaces reduce the IIS strength by improving the charge screening, and this boosts the PF.

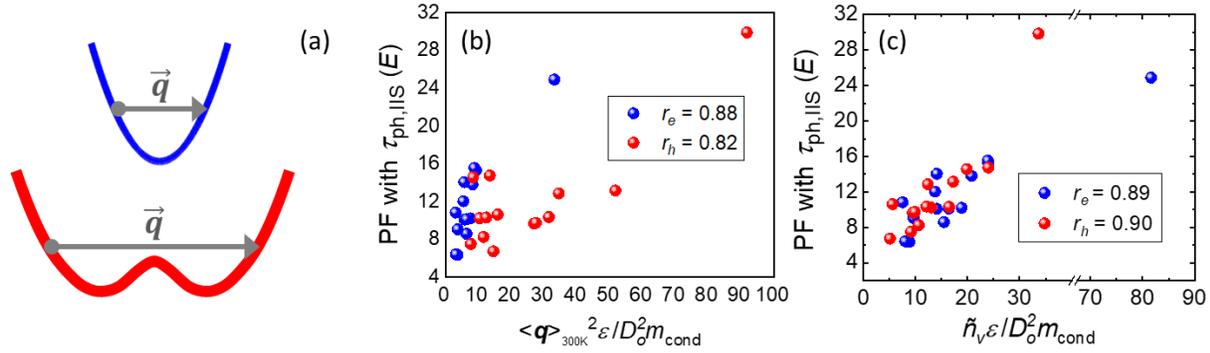

**Figure 7.** (a) Schematic of a regular parabolic band, in blue, and of a warped band with two (or in general many) valleys, in red, with the ionized impurity scattering vector $\vec{q}$ indicated. In the case of warped, irregular bands, with many valleys, $\vec{q}$ can be much larger (at the same energy), increasing the carrier screening effectiveness and decreasing the IIS scattering rate. (b) The averaged $\langle q \rangle^2$ is added to the descriptor of Figure 6e, together with the dielectric constant $\varepsilon$ to produce a generic descriptor that can be employed in different materials situations. The Pearson correlation for the hole transport cases (note the valence bands are strongly warped) is improved. (c) The average number of valleys per band counted in the 0.1 eV range from the band edge is used as an easily accessible parameter to mimic the surface width and average IIS scattering exchange vector. The Pearson correlation is improved compared to Figure 6e as well.

The correlation between the PF and the upgraded descriptor $\widetilde{n_v}\varepsilon_r / D_o^2 m_{cond}$, is shown in **Figure 7c**, where $\widetilde{n_v}$ is the average number of valleys per band. The descriptor is now very satisfactory for both *n*-type and *p*-type materials ($r_e$ = 0.89 and $r_h$ = 0.90) and can be easily used instead of the scattering vector. Note that including the number of valleys will make the descriptor even stronger in the general case where inter-valley scattering is suppressed, independent of the scattering mechanism. However, in the half-Heusler materials we consider the inelastic optical phonons are dominant and inter-valley/inter-band scattering could be significant. The average number of valleys per band, $\widetilde{n_v}$, however, can be changed to the total number of valleys $n_v$ as $n_v \varepsilon_r / D_0^2 m_{\text{cond}}$ with almost identical correlations outcomes (which reflects to the validity of the descriptor earlier in Figure 7 as well). Importantly, however, it is understood that the number of valleys is essential in cases where inter-valley scattering is suppressed for the dominant scattering mechanisms. In materials with wide constant energy surfaces, and under IIS dominant transport (as in the optimal cases for TE materials), due to enhanced screening we are in the regime of inter-valley suppression, and the use of $n_v$ is imperative. We underline that in NbCoSn, one of the best *p*-type materials we computed, the valence band has the highest number of valleys amongst the materials shortlist considered. In particular, the valence bands has a valley at the W point, which is 6-fold degenerate (24 W points shared between 4 adjacent zones) and relatively far from the other high-symmetry points (except from the X point). Thus having valleys at W is very efficient in suppressing inter-valley scattering, especially when there are no valleys in X. This supports recent findings [62] that suggest that having valleys in W is helpful in gaining high PF.

### 3.2.3 Validity of the new descriptors at higher temperatures

The analysis up to now considered only room temperature, *T* = 300 K. It is necessary to evaluate the performance of the descriptors for higher temperatures, however, since the majority of TE materials are more efficient at higher temperatures. **Figure 8** shows the Pearson correlation coefficients for the maximum PFs of the materials under investigation for $T = 300$ K, 600 K, and 900 K. Several combinations of the parameters that compose the final descriptor are separately analyzed. The intent is to highlight which has the most dominant role for electrons (nearly parabolic bands) in **Figure 8a** and **Figure 8b**, and holes (complex multiple-valleys bands) in **Figure 8c** and **Figure 8d**. In **Figure 8a** and **Figure 8c** (first column) we show results for phonon-limited transport, and in **Figure 8b** and **Figure 8d** for phonon plus IIS transport conditions. The conductivity effective mass $m_{\text{cond}}$ is noted as $m_c$ in the figures for brevity.

In the phonon-limited case, for *n*-type materials with (nearly) parabolic bands (**Figure 8a**), at 300 K the descriptor $1/D_o^2$ performs well and is improved with increasing temperature (red-dotted line). On the other hand, although the performance of $1/m_c$ is better at 300 K compared to $1/D_o^2$, with $r_e = 0.74$ (black-squared line), its performance decreases significantly with temperature. This is possibly because at higher temperatures the non-parabolicity and warping of the bands at higher energies become more relevant. For *p*-type materials with large and warped constant energy surfaces, the $1/m_c$ is not a good descriptor at any temperature with $r_h < 0.1$ (**Figure 8c**), possibly for the same reason as for *n*-type at high temperatures. The $1/D_o^2$ is a strong parameter for *p*-type phonon-limited transport, and in both carrier-type cases the combination of $1/D_o^2 m_c$ is the dominant one with $r_e = 0.96$ and $r_h = 0.95$ (green-triangle lines).

In the phonon plus IIS transport case for *n*-type nearly parabolic bands (Figure 8b), the $1/D_o^2$ descriptor (red dots) provides a fair correlation, which improves with temperatures. When enriched with $1/m_c$, the $1/D_o^2 m_c$ descriptor performs well across temperatures with $r_e = 0.86$ at 300 K and $r_e = 0.83$ at 900 K (green-triangle line). A further improvement occurs at 300 K by incorporating $\varepsilon_r$ and the number of valleys, $n_v$, leading to $n_v \varepsilon_r / D_o^2 m_{\text{cond}}$ with $r_e = 0.90$ (blue pentagons). Its strength weakens at medium temperatures. Still, however, $n_v \varepsilon_r / D_o^2 m_{cond}$ is retained at acceptable levels $r_e = 0.84$ at 600 K and 0.78 at 900 K.

The $1/D_o^2 m_c$ descriptor retains its strength for the warped highly non-parabolic *p*-type bands (Figure 8d) with $r_h = 0.78$ at 300 K and $r_h = 0.94$ at 900 K (green triangles). This originates from the synergy between $1/m_c$ and $1/D_o^2$, which individually perform very poorly. In this case, an additional benefit to the descriptor is the addition of the average exchange vector $<q>^2$, which improves the $r_h$ to 0.82 at $T = 300$ K up to 0.94 when $T = 900$ K. This signifies the importance of suppressing intervalley scattering in the presence of the dominant IIS. Note that this is important in the *p*-type materials we investigate, as their bandstructures consist of large and warped constant energy surfaces. The exchange vector can be effectively substituted by the band-averaged number of valley (see the Supporting Information for details). However, when we use the total number of valleys (blue pentagon) the descriptor loses its performance because of the low performance of $n_v$ alone (light blue diamonds).

This signals that more than the number of valleys itself, it is their distance in the BZ which matters. Thus, many valleys placed far away from each other in the BZ signify a favourable situation, whereas many valleys in the same or nearby BZ points represents a non-favourable situation. Moreover, the DOS effective mass of the materials with higher number of valleys increases with temperature, signalling that when inter-valley scattering is strong, it takes away possible positive effect of high degeneracy. It turns out, however, that the $n_v \varepsilon_r / D_o^2 m_{cond}$ descriptor (blue pentagons) is the best trade-off between simplicity and accuracy also for the *p*-type case, with $r_h > 0.75$ over the whole temperature range. Thus, $n_v \varepsilon_r / D_o^2 m_{cond}$ (blue pentagon lines) is the only descriptor that provides good to very high correlation results for most cases, across carrier type, scattering mechanisms, and temperatures.

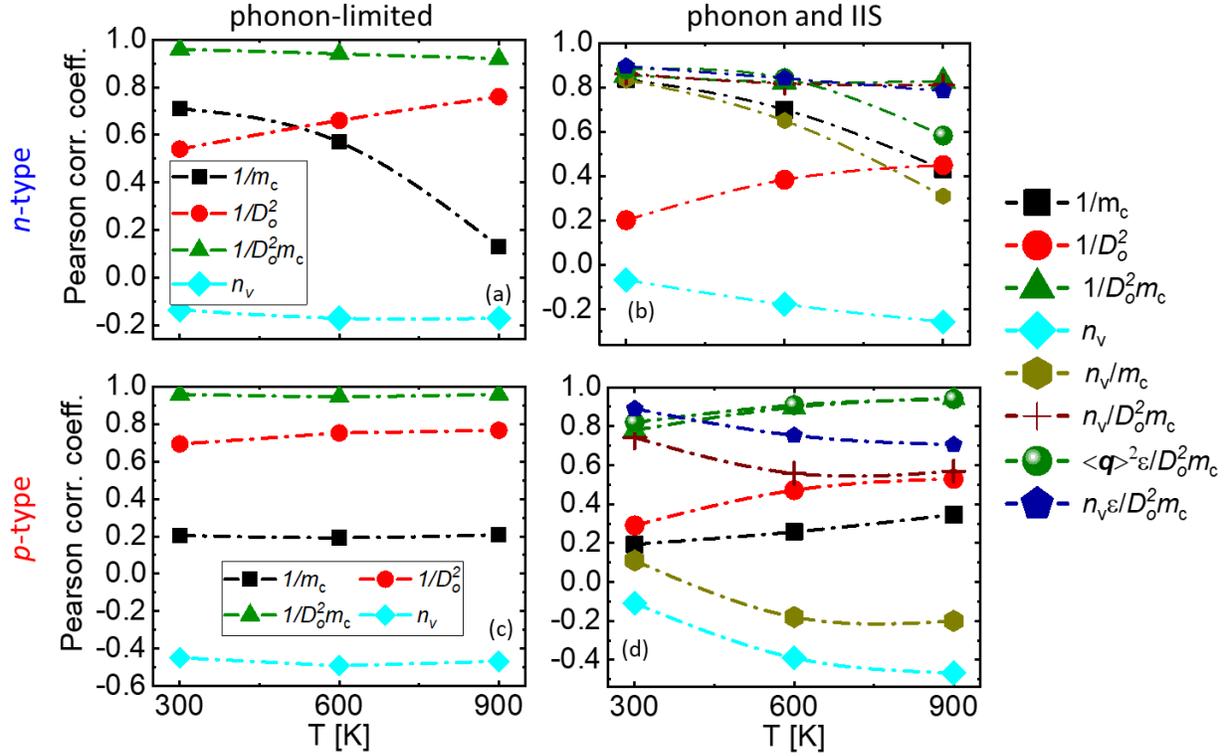

**Figure 8.** Power factor (PF) peak value Pearson correlations coefficients with multiple descriptors, computed at temperatures of 300 K, 600 K, and 900 K in the case of phonon-limited transport (first column) and phonon plus IIS transport (second column). The first row, (a-b), shows results for *n*-type materials and the second row, (c-d), for *p*-type materials.

### 3.3 Bipolar transport

The bandgap is historically one of the most relevant features to signal TE material performance.[11] However, the direct consideration of the energy gap size in the ranking of TE materials does not lead to straightforward conclusions.[2] The compounds we consider have bandgaps ranging from 0.19 eV to 1.14 eV, which makes it necessary to consider the intrinsic thermally generated carriers and the minority carriers as well, i.e. bipolar transport effects, especially at elevated temperatures.

In this case, the transport calculation should take into account both conduction and valence bands combined, and this requires a slight modification in the descriptors, as we show below.

A simulation difference between bipolar and unipolar transport is the different shift of the Fermi level with temperature, which influences the transport calculations significantly. In the unipolar case, as the temperature increases, the Fermi level shifts down (for $n$-type, for example), in order for charge neutrality to be satisfied, such that the mobile carrier density equals the doping density. In the bipolar case, for charge neutrality we need to consider also the intrinsic thermally generated carriers and the minority carriers as well, whose concentration increases with temperature.[63] The increased density of both types of carriers leads to a reduced Fermi level shift compared to the unipolar case. Importantly, in this case both electrons and holes contribute to the screening of ionized impurities in an additive fashion.[34, 64]

The correlation between the peak PF computed under unipolar and bipolar transport considerations for the phonon plus IIS case is shown in **Figure 9a** at $T = 900$ K where bipolar effects are relevant. The dashed line in the figure has a slope of unity and serves as guide-for-the-eye. The peak PF values feature an extremely high correlation, $r > 0.9$. At the lower $T = 300$ K the correlation is complete (not shown). It is also almost perfect in the case of 900 K and phonon-limited scattering. The magnitude of the PF, is of course reduced in the low bandgap materials, however, this does not affect the materials ranking, except for ScNiBi, which has the smallest bandgap amongst the investigated compounds. This is because in the materials we examine, the change in the PF due to bipolar effects is not larger in general compared to the differences in the PF between materials.

In **Figure 9b** and **Figure 9c** we show the Pearson correlations for several descriptors at different temperatures for $n$-type and $p$-type materials, respectively. It is clear that the same descriptors that provide optimal correlations for the unipolar case, are also the most relevant ones for the bipolar case, i.e. the ones involving $1/D_o^2 m_{\text{cond}}$ for *n*-type ($r_e > 0.6$) and the ones involving $n_v$ for *p*-type ($r_h > 0.7$). The temperature trends are also similar in several cases. The descriptor developed earlier, which provided high performance for both carrier types, $\langle q \rangle^2 \varepsilon_r / D_o^2 m_{\text{cond}}$ still performs exceptionally well for the *p*-type case, with Pearson correlation coefficients of $r_h \sim 0.85$ at high temperatures. For *n*-type, however, it drops to $r_e \sim 0.45$ at 900 K. The observation is similar for the $n_v \varepsilon_r / D_o^2 m_{\text{cond}}$ descriptor, which results in Pearson correlation coefficients of $r_h = 0.73$, and $r_e = 0.82$ at 900 K. Thus, for bipolar transport in the more realistic case of phonon plus IIS transport, a modification to this descriptor is not needed. However, a simple modification is generally proposed in the literature by using the square of the reduced band gap, $\widetilde{E_G} = \frac{E_G}{k_B T}$, to capture the reduction to the Seebeck coefficient with the band gap (squared as the Seebeck appears squared in the PF).[11] The descriptor then becomes $\widetilde{E_G}^2 n_v \varepsilon_r / D_o^2 m_{\text{cond}}$, but its Pearson correlation values at 900 K drop to $r_e = 0.21$ for *n*-type and $r_h = -0.43$ for *p*-type, as indicated by the center-dotted circles shown in Figure 9b,c, respectively. The inclusion of the band gap in this latter descriptor does not lead to a straightforward improvement, as already indicated in the literature,[2] and the common consideration of it [11] requires further studies.

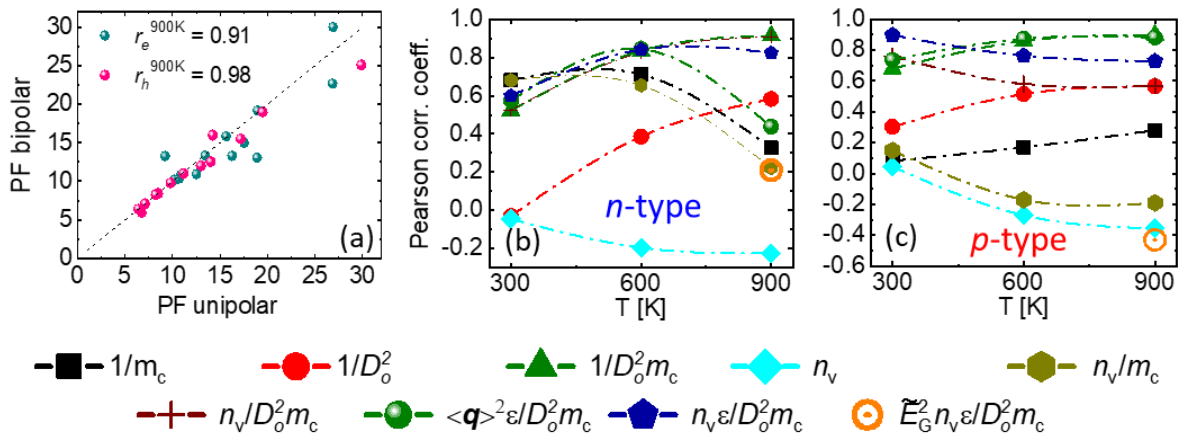

**Figure 9.** (a) Correlation between power factor (PF) peak values computed under bipolar and unipolar transport considerations for phonon plus IIS transport at 900 K. (b) and (c) PF peak value Pearson correlation coefficients with multiple descriptors, computed at 300 K, 600 K, and 900 K in the case of phonon plus IIS transport for *n*-type and *p*-type materials, respectively.

### 3.4 Discussion

We have identified promising material candidates in section 3.1, both *n*-type and *p*-type. For *n*-type the more promising material candidates are NbFeSb, HfNiSn, YNiBi and ScNiBi with peak PFs > 15 mW/(m·K$^2$). Other than NbCoSn, these are usually *p*-type TE materials so the introduction of donor impurities could present a materials science challenge. The most promising *p*-type materials are ScNiBi, NbCoSn, NbFeSb and TiNiSn with PFs ≥ 12 mW/(m·K$^2$). For the former, the use of acceptor dopants has been demonstrated.[39] The latter is poorly studied, but it is in principle *p*-type dopable with Sn in the Bi lattice site.[52] The descriptors developed indicate why these compounds are the more promising among the ones we have studied. They have a good combination of a low optical deformation potential, a low conductivity effective mass, and a high dielectric constant. In the case of charge transport in the warped, multi-valley, valence bands, their many valleys for transport also result in large constant energy surfaces with large scattering vectors, which improve the charge screening and increase the conductivity.

The NbCoSn is interesting as it performs quite well for both *n*- and *p*-type. Both can be experimentally realized by substituting Sn with Sb or Bi for *n*-type, and by substituting Nb with Ti or Zr and Co with Fe for *p*-type.[39, 40, 51] This would enable a TEG with the *n*-type and *p*-type legs from the same material, which could improve the TEG's mechanical stability and life endurance.[49]

We have shown the importance of considering electron scattering with ionized dopants in power factor studies, and proposed the $n_v \varepsilon_r / D_0^2 m_{\text{cond}}$ descriptor. Low deformation potentials, low conductivity effective masses, large constant energy surfaces through many valleys placed far from each other in the reciprocal space, and high dielectric constants, can compose a material properties 'shopping list' (experimentally and computationally). While the dielectric constant is usually computed in *ab-initio* calculations and the conductivity effective mass and number of valleys can be easily extracted from the 3D bandstructure, the deformation potentials are rarely computed in first principle calculations. For example, we register only two, very recent attempts to compute from first principles the deformation potentials in half-Heusler alloys.[9, 24, 25, 65] The situation for other TE material families is similar.[66, 67] We suggest that in the predictive search for better thermoelectrics, deformation potential calculations are also performed – at least for regular nearly parabolic bandstructure material.

It is worth to underline the importance of the deformation potential method. In the deformation potential theory, the electron-phonon coupling is described by a single parameter, the deformation potential. This theory can have some limitations in respect of a precise evaluation of the coupling strength for each possible pair of initial and final state and each possible phonon, especially for complex band materials. Nevertheless, the latter treatment is computationally extremely expensive and, at present, can be successfully used for a relatively fast and automated computation of the transport properties only in 2D materials and with the restriction to certain portions of the BZ, thanks to the reduced number of employed *k*-points.[20] The use of deformation potentials can be a significant step forward compared to constant relaxation times. Besides, the proper and feasible computation of the deformation potential values can be extremely useful in the materials ranking and pre-screening. Indeed, the predictive evaluation of the deformation potentials, without available experimental data for comparison, is an on-going effort in the thermoelectric community. [9, 65–67]

Finally, we note that there is significant effort undertaken in order to achieve band alignment in complex bandstructure materials for high power factors. Recent works show that this is beneficial only in the case where the bands brought closer to the band edge are lighter, and when inter-valley scattering is weak. [27, 68] This is very well aligned with what our descriptors suggest, i.e. sharp bands with light conductivity effective masses and large energy surfaces with strong charge screening, which effectively leads to weakened inter-valley IIS scattering. Thus, the descriptors we developed can also be applicable in band alignment studies for improving the thermoelectric power factor.

## 4. Conclusions

In summary, we have developed a computational method that combines energy dependent electron-phonon and electron-ionized impurity scattering mechanisms with DFT electronic structures. We then investigated the thermoelectric power factor performance of a group of 14 half-Heuslers, for both *n*-type and *p*-type transport. The comparison of the materials' performances under the constant relaxation time approximation and energy dependent scattering assumptions shows absence of any correlation in the materials ranking. We tested the validity of a set of descriptors currently used in the literature for materials screening exercises, and have developed new, highly improved descriptors for the thermoelectric power factor, based on our detailed scattering treatment. Our proposed descriptor $n_v \varepsilon_r / D_o^2 m_{\text{cond}}$ is based on number of valleys, the dielectric constant, the conductivity effective mass, and the deformation potential for the dominant electron-phonon process. In the case of strong bipolar effects, it can be modified with the inclusion of the reduced bandgap to $\widetilde{E_G}^2 n_v \varepsilon_r / D_o^2 m_{\text{cond}}$.

We strongly suggest that every thermoelectric materials screening study should consider in more detail: i) the energy and wave-vector dependence of the scattering processes, ii) the ionized impurity scattering, and iii) the volume that the bands occupy in the Brillouin Zone. We further show, in the case of *n*-type half-Heuslers, which have nearly parabolic bands, a highly relevant and simplest descriptor is $1/D_o^2 m_{\text{cond}}$. For warped non-parabolic multi-band/valley bandstructures, as the *p*-type half-Heuslers, the most important transport aspect is the ability of the carriers to screen the dopant impurities, which leads to inter-valley scattering suppression. Only then, the different valleys can contribute to transport independently and improve the PF. The number of valleys, $n_v$, can then be used as a single parameter descriptor, as it i) reflects the average size of the constant energy surface, and the larger it is, the stronger the IIS screening is, which allows the valleys to ii) independently contribute to transport without increasing scattering for carriers from other valleys. The insight gained from this work can serve as guidance to understand experimental observations. The proposed descriptors can be useful materials screening tools to advance the search for high performance thermoelectric materials.

ASSOCIATED CONTENT

**Supporting Information**. Tables with scattering input parameters and extracted effective masses, procedure to extract the effective masses. This material is available free of charge via the Internet at http://pubs.acs.org.

AUTHOR INFORMATION

**Corresponding Author**

*Patrizio.Graziosi@warwick.ac.uk, Patrizio.Graziosi@gmail.com

**Author Contributions**

The manuscript was written through contributions of all authors.

**Funding Sources**

European Commission under the Grant agreement ID: 788465 (GENESIS - Generic semiclassical transport simulator for new generation thermoelectric materials) and European Research Council (ERC) under the European Union's Horizon 2020 Research and Innovation Programme (Grant Agreement No. 678763).

ACKNOWLEDGMENT

This work has received funding from the Marie Skłodowska-Curie Actions under the Grant agreement ID: 788465 (GENESIS - Generic semiclassical transport simulator for new generation thermoelectric materials) and from the European Research Council (ERC) under the European Union's Horizon 2020 Research and Innovation Programme (Grant Agreement No. 678763). The authors thank Laura De Sousa Oliveira, Dhritiman Chakraborty, Alberto Riminucci and Alek Dediu for reading the manuscript.

# Supporting Information

## Material descriptors for the discovery of efficient thermoelectrics

*Patrizio Graziosi\*, Chathurangi Kumarasinghe, Neophytos Neophytou*

\* patrizio.graziosi@cnr.it

**Table S1.** The scattering parameters as used in the simulations presented in this work, extracted from the literature.[9, 32, 33]

| Material | $\rho$ [kg/m³ ×10³] | $u_s$ [m/s ×10³] | ADP [a)] [eV] | ODP [a)] [eV/m ×10¹⁰] | $\hbar\omega$ [eV] | $\varepsilon_r$ [$\varepsilon_0$] |
|---|---|---|---|---|---|---|
| HfCoSb | 9.54 | 5.64 | 0.2 / 0.4 | 1.4 / 1.85 | 0.028 | 17.51 |
| HfNiSn | 10.3 | 3.39 | 0.1 / 0.7 | 1.8 / 2 | 0.028 | 20.88 |
| NbCoSn | 8.43 | 5.36 | 0.2 / 0.5 | 2.6 / 2.16 | 0.034 | 22.6 |
| NbFeSb | 8.45 | 3.99 | 1 / 0.8 | 1.6 / 2.1 | 0.036 | 22.99 |
| ScNiBi | 8.48 | 3.1 | 0.8 / 0.2 | 2.2 / 1.4 | 0.028 | 29 |
| ScNiSb | 6.6 | 3.9 | 0.7 / 0.4 | 2.3 / 1.8 | 0.032 | 19.66 |
| TiCoSb | 7.4 | 4.04 | 1 / 0.5 | 1.75 / 2.2 | 0.036 | 19.09 |
| TiNiSn | 7.2 | 3.9 | 0.7 / 0.2 | 2.3 / 0.9 | 0.034 | 22.92 |
| YNiBi | 8.6 | 2.9 | 0.5 / 0.9 | 2.3 / 2.7 | 0.022 | 26.55 |
| YNiSb | 6.9 | 3.5 | 0.4 / 0.9 | 2.35 / 3.14 | 0.026 | 19.74 |
| ZrCoBi | 9.83 | 3.2 | 0.2 / 0.8 | 2.1 / 1.8 | 0.028 | 20.37 |
| ZrCoSb | 7.14 | 5.55 | 0.2 / 1 | 2.05 / 1.75 | 0.028 | 17.87 |
| ZrNiPb | 9.6 | 3.39 | 0.36 / 0.60 | 2.8 / 1.6 | 0.024 | 23.6 |
| ZrNiSn | 7.62 | 3.97 | 0.35 / 0.3 | 2.8 / 1.8 | 0.029 | 20.9 |

[a)] electrons / holes

**Table S2.** Extracted bandstructure parameters for *n*-type materials. The effective masses are in units of the electron rest mass and the total number of valleys considers the valleys within 0.1 eV from the band edge. $n_b$ is the number of bands and $\widetilde{n_v}$ the average number of valleys per band.

| Material | $m_{DOS}$ $T = 300$ K | $m_{DOS}$ $T = 600$ K | $m_{DOS}$ $T = 900$ K | $m_{cond}$ $T = 300$ K | $m_{cond}$ $T = 600$ K | $m_{cond}$ $T = 900$ K | $n_v$ | $n_b$ | $\widetilde{n_v}$ |
|---|---|---|---|---|---|---|---|---|---|
| HfCoSb | 4.31 | 3.99 | 3.8 | 1.42 | 1.33 | 1.26 | 3 | 1 | 3 |
| HfNiSn | 4.1 | 3.5 | 3.29 | 0.93 | 0.96 | 1.0 | 3 | 1 | 3 |
| NbCoSn | 3.64 | 3.68 | 3.8 | 1.33 | 1.29 | 1.32 | 6 | 2 | 3 |
| NbFeSb | 0.92 | 0.88 | 0.9 | 0.33 | 0.37 | 0.61 | 3 | 1 | 3 |
| ScNiBi | 3.12 | 2.83 | 2.69 | 0.75 | 0.73 | 0.72 | 3 | 1 | 3 |
| ScNiSb | 3.02 | 2.78 | 2.65 | 0.79 | 0.78 | 0.77 | 3 | 1 | 3 |
| TiCoSb | 4.97 | 4.92 | 5.05 | 2.12 | 2.03 | 2.03 | 3 | 1 | 3 |
| TiNiSn | 2.43 | 2.4 | 2.46 | 0.84 | 0.89 | 0.94 | 3 | 1 | 3 |
| YNiBi | 3.26 | 2.79 | 2.57 | 0.63 | 0.61 | 0.6 | 3 | 1 | 3 |
| YNiSb | 3.45 | 3.03 | 2.82 | 0.76 | 0.73 | 0.71 | 3 | 1 | 3 |
| ZrCoBi | 4.49 | 4.13 | 3.94 | 1.44 | 1.34 | 1.28 | 3 | 1 | 3 |
| ZrCoSb | 4.36 | 4.07 | 3.9 | 1.59 | 1.49 | 1.42 | 3 | 1 | 3 |
| ZrNiPb | 1.93 | 1.86 | 1.84 | 0.55 | 0.57 | 0.58 | 3 | 1 | 3 |
| ZrNiSn | 1.94 | 1.87 | 1.85 | 0.58 | 0.6 | 0.61 | 3 | 1 | 3 |

**Table S3.** Extracted bandstructure parameter for *p*-type materials. The effective masses are in units of the electron rest mass and the total number of valleys considers the valleys within 0.1 eV from the band edge. $n_b$ is the number of bands and $\widetilde{n_v}$ the average number of valleys per band.

| Material | $m_{DOS}$ $T = 300$ K | $m_{DOS}$ $T = 600$ K | $m_{DOS}$ $T = 900$ K | $m_{cond}$ $T = 300$ K | $m_{cond}$ $T = 600$ K | $m_{cond}$ $T = 900$ K | $n_v$ | $n_b$ | $\widetilde{n_v}$ |
|---|---|---|---|---|---|---|---|---|---|
| HfCoSb | 7.8 | 8.31 | 8.58 | 1.44 | 1.55 | 1.73 | 11 | 3 | 3.67 |
| HfNiSn | 1.88 | 1.8 | 1.78 | 0.55 | 0.56 | 0.57 | 3 | 3 | 1 |
| NbCoSn | 5.79 | 6.67 | 7.2 | 1.08 | 1.2 | 1.34 | 16 | 3 | 5.33 |
| NbFeSb | 5.07 | 4.84 | 4.85 | 1.05 | 1.08 | 1.13 | 8 | 2 | 4 |
| ScNiBi | 1.92 | 1.71 | 1.64 | 0.44 | 0.49 | 0.52 | 3 | 3 | 1 |
| ScNiSb | 2.2 | 1.96 | 1.87 | 0.49 | 0.55 | 0.58 | 3 | 3 | 1 |
| TiCoSb | 7.35 | 8.6 | 9.16 | 1.58 | 1.88 | 2.14 | 11 | 3 | 3.67 |
| TiNiSn | 6.11 | 5.33 | 5.11 | 1.64 | 1.64 | 1.69 | 3 | 3 | 1 |
| YNiBi | 1.43 | 1.28 | 1.25 | 0.3 | 0.36 | 0.41 | 3 | 3 | 1 |
| YNiSb | 1.79 | 1.62 | 1.59 | 0.39 | 0.46 | 0.51 | 3 | 3 | 1 |
| ZrCoBi | 6.59 | 6.74 | 7.09 | 1.4 | 1.45 | 1.58 | 11 | 3 | 3.67 |
| ZrCoSb | 6.63 | 6.63 | 6.96 | 1.46 | 1.54 | 1.7 | 8 | 3 | 2.67 |
| ZrNiSn | 3.33 | 2.97 | 2.88 | 0.93 | 0.95 | 1.01 | 3 | 3 | 1 |
| ZrNiPb | 4.49 | 3.98 | 3.94 | 1.15 | 1.2 | 1.36 | 3 | 3 | 1 |

**Extraction of the density of states and conductivity effective masses from bandstructure**

The data analysis and proposed descriptors make use of the DOS and conductivity effective masses $m_{DOS}$ and $m_{cond}$, respectively. These are effective quantities that include information from the entire bandstructure, rather just specific valleys. To extract these, we consider non-degenerate conditions, where the Fermi level is placed in the bandgap, several $k_BT$ from the bands. Then the $m_{DOS}$ is the effective mass of an isotropic parabolic band that provides the same carrier density as the actual bandstructure. The $m_{cond}$ is the effective mass of an isotropic parabolic band that has the same average velocity as all that of the entire bandstructure.[69]

To compute $m_{DOS}$ we utilize the common expression for the carrier density in a non-degenerate semiconductor:

$$N = N_c e^{-\frac{E_c - E_F}{k_B T}} \quad (S1)$$

where $E_c$ is the band edge, $E_F$ is the Fermi level, $k_B$ is the Boltzmann constant, $T$ is the temperature and $N_c = 2\left(\frac{m_{DOS}k_B T}{2\pi\hbar^2}\right)^{3/2}$ is the effective density of states in the conduction band. This expression is derived by approximating the Fermi Dirac statistics with Boltzmann statistics. The carrier density in the numerical DFT bandstructure is given by:

$$N = \frac{2}{(2\pi)^3}\sum_{\boldsymbol{k},n} f_{E_{\boldsymbol{k},n}} dV_{\boldsymbol{k}} \quad (S2)$$

where the sum runs over all the $\boldsymbol{k}$ points in the reciprocal unit cell (mesh used for the DFT calculations) and all the bands, $f_{E_{\boldsymbol{k},n}}$ is the Fermi-Dirac distribution evaluated at the energy of each state (approximated by Boltzmann statistics in the non-degenerate limit), and $dV_{\boldsymbol{k}}$ is the volume element in the $\boldsymbol{k}$ space that depends only on the mesh. By equating the carrier density from equations (S1) and (S2) for the same Fermi level and band edge, we can extract the effective density of states in the conduction band, $m_{DOS}$. The process is the same for the valence band.

The conductivity effective mass of the bandstructure, $m_{cond}$, is extracted from the so-called injection velocity $v_{inj}$ of the carriers in non-degenerate conditions.[69] As this method is derived for MOSFET devices,[70] it considers the ballistic current that the positive velocity states of the bandstructure will allow for a specific Fermi level in the non-degenerate conditions as:

$$I_{+,n} = e \frac{1}{2} \frac{2}{(2\pi)^3} \sum_{k_n} f_{(E_{k_n} - E_{F,S})} |v_{k_n}| dV_{k_n} \quad (S3)$$

where $e$ is the electron charge, the ½ factor is to take into account only half of the states, and, $|v_{k_n}|$ is the band carrier velocity in absolute terms in order to only account for positive velocities. Then, the injection velocity is extracted by dividing the positive going ballistic current with the carrier density (to get the average velocity per carrier):

$$v_{\text{inj}} = \frac{I_{+,n}}{e \frac{1}{2} \frac{2}{(2\pi)^3} \sum_{k_n} f_{(E_{k_n} - E_{F,S})} dV_{k_n}} \quad (S4)$$

where the denominator in equation (S4) is the charge occupying the positive velocity states alone. Once the injection velocity is extracted, the conductivity effective mass can be computed as:

$$m_{\text{cond}} = \frac{2k_B T}{v_{\text{inj}}^2 \pi} \quad (S5)$$

which is the corresponding quantity for a simple parabolic band.[69, 70]

This calculation is performed along the three Cartesian directions *x*, *y*, and *z*. Thus, three conductivity affective masses are extracted, $m_{\text{cond},x}$, $m_{\text{cond},y}$, and $m_{\text{cond},z}$ and a final average conductivity affective mass is then formed by $m_{\text{cond}} = \frac{3}{m_{\text{cond},x}^{-1} + m_{\text{cond},y}^{-1} + m_{\text{cond},z}^{-1}}$. The process is performed twice, once for the conduction and once for the valence bands.

As an example, we show in **Figure S1a** the numerically extracted DOS for HfCoSb (blue lines) and the DOS provided by the extracted $m_{\text{DOS}}$, assuming that it forms an isotropic, parabolic band (dashed-dot green line), indicating adequate match at low energies. In **Figure S1b** we show the HfCoSb valence band injection velocity using the numerical bandstructure (blue line), and the injection velocity for an isotropic parabolic band with the extracted $m_{\text{cond}}$ value (dash-dot green line), again indicating adequate match in the low-density region.

The relevance of $m_{\text{cond}}$ becomes clear when comparing the TDF computed for the actual numerical bandstructure and for an isotropic parabolic band of an effective mass equal to the extracted $m_{\text{cond}}$, **Figure S1c**. The blue solid line (numerical bandstructure) and the green dashed-dot line (parabolic band with $m_{cond}$) are for the ADP-limited scattering mechanism. The lines overlap for a large energy region, indicating that $m_{cond}$ captures the core transport information for ADP scattering in complex bandstructures. In the ODP case, the TDFs for the solid orange line (for the numerical bands) and the purple dash-dot line (for the parabolic band of $m_{cond}$) also clearly overlap. The TDFs deviate far from the band onset, when the parabolicity is lost.

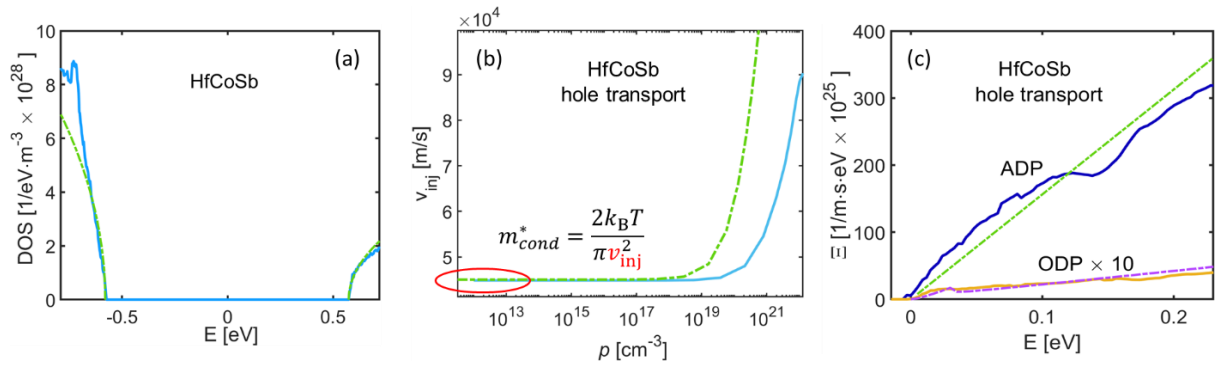

**Figure S1:** Comparisons between quantities extracted from the full numerical bandstructure and the corresponding quantities computed from the extracted $m_{DOS}$ and $m_{cond}$ for HfCoSb. (a) The DOS of HfCoSb, blue line, and the DOS of parabolic bands with the isotropic $m_{DOS}$ values, green dashed lines. (b) Injection velocity versus carrier concentration for the positive going carriers of the HfCoSb valence band under ballistic conditions using the full bandstructure, blue solid line, and the parabolic isotropic band with the extracted effective mass value $m_{cond}$, green dash-dotted line. (c) Transport distribution function versus energy for the HfCoSb valence band using the full bandstructure and the band formed with the extracted $m_{cond}$. The case of acoustic phonons (ADP – Acoustic Deformation Potential) transport and optical phonons (ODP – Optical Deformation Potential), transport are plotted. The blue and yellow solid lines indicate the full bandstructure TDFs. The green and purple dash-dotted lines indicate the TDFs for the parabolic isotropic band having an effective mass of $m_{cond}$.